\documentclass{mem}
\usepackage{natbib}\usepackage{txfonts}\usepackage{balance}
\usepackage{graphicx}
\idline{77}{1}
\begin{document}

%% START user-defined macros

\def\aabun{[$\alpha$/Fe]}
\def\av{A$_{\rm V}$}
\def\teff{${\rm T}_{\rm eff}$}
\def\logg{$\log g$}
\def\feh{[Fe/H]}
\def\age{$\tau$}
\def\parallax{$\pi$}

\def\deg{$^{\circ}$}
\def\micron{$\mu$m}
\def\uas{$\mu$as}

\def\Msol{M$_{\odot}$}
\def\Rsol{R$_{\odot}$}
\def\Lsol{L$_{\odot}$}

%% END user-defined macros

\title{
Prospects for Gaia and other space-based surveys
}

\subtitle{}

\author{
Coryn A.L.\ Bailer-Jones
%\inst{1,2}
}

%\offprints{}

\institute{
Max-Planck-Institut f\"ur Astronomie, K\"onigstuhl 17,
69117 Heidelberg, Germany\\
\email{calj@mpia.de}
}

\authorrunning{Bailer-Jones}

\titlerunning{Prospects for Gaia and other space-based surveys}

\abstract{ Gaia is a fully-approved all-sky astrometric and photometric survey
  due for launch in 2011. It will measure accurate parallaxes and proper
  motions for everything brighter than G=20 (ca.\ $10^9$ stars). Its primary
  objective is to study the composition, origin and evolution of our Galaxy
  from the 3D structure, 3D velocities, abundances and ages of its stars.  In
  some respects it can be considered as a cosmological survey at redshift
  zero. Several other upcoming space-based surveys, in particular JWST and
  Herschel, will study star and galaxy formation in the early (high-redshift)
  universe.  In this paper I briefly describe these missions, as well as SIM
  and Jasmine, and explain why they need to observe from space. I then discuss
  some Galactic science contributions of Gaia concerning dark matter, the
  search for substructure, stellar populations and the mass--luminosity
  relation.  The Gaia data are complex and require the development of novel
  analysis methods; here I summarize the principle of the astrometric
  processing.  In the last two sections I outline how the Gaia data can be
  exploited in connection with other observational and theoretical work in
  order to build up a more comprehensive picture of galactic evolution.
  \keywords{Space vehicles -- Methods: data analysis -- Surveys -- Astrometry -- Galaxy:
  general} 

\vspace*{2ex}
\begin{center}
{\em Proceedings of JD13, IAU General Assembly 2006}
\end{center}
\vspace*{-1ex}
}
  \maketitle{}

\section{Why observe from space?}

Observing from space is not merely desirable. It is {\em essential} for
certain types of observation or instrumentation. The Earth's atmosphere
strongly absorbs over much of the electromagnetic spectrum, in particular in
the X-ray, UV and near- and far-infrared. Most of these regions cannot be
observed from the ground at all. Even in regions which are accessible
from the ground (such as various near-infrared windows), the background in
space is much fainter, permitting deeper observations. Above the
atmosphere we can also achieve diffraction-limited imaging. As well as
increasing the spatial resolution, this has the additional advantage that 
deeper observations of point sources are possible (because less background
light is integrated into the smaller point-spread function). Temporal
and spatial variations in the refractivity of the Earth's atmosphere limit the
accuracy of wide-field astrometry to about 1\,mas. It is primarily for this
reason that all accurate optical wide-field astrometry projects
must be performed from space.

Space offers a stable environment in a more general sense. Weather and the
diurnal and annual cycles on the Earth give rise to significant temperature
variations, creating significant problems in accurate metrology (required in
astrometry or interferometry). To avoid these disturbances, instruments have to
be placed in large (and expensive) isolating vessels and/or the variations
must be carefully monitored and calibrated. Human and seismic activity adds
further mechanical disturbances. Various orbits in space offer an environment
which is mechanically and thermally much more stable. Suitable orbits include
the L1, L4 and L5 Lagrange points of the Earth--Sun system. (L3 is not visible
from the Earth so is of limited use.) The L2 point actually sits in the shadow
of the Earth, and is dynamically unstable (as are L1 and L3) but Lissajous
orbits about this point is stable for extended periods and can avoid eclipses
(by the Earth, but not necessarily the Moon) for many years.  These orbits are
far from any residual Earth atmospheric drag and so offer very stable
environments.  They are attractive for astrometric and interferometric
instruments which require stability or control at levels which are difficult
to achieve on the ground.  Herschel, Planck, Gaia and
JWST will all be placed in L2 orbits (WMAP is already in one) and LISA
Pathfinder will be (and SOHO already is) in an L1 orbit. Other orbits offer
similar conditions, such as an Earth-trailing heliocentric one (used by
Spitzer and planned for SIM).

A space telescope can access the entire sky over a short time period.
This is important for global astrometry (see section~\ref{gaia_dp}).
Finally, space allows multiple satellites to be manoeuvred into arbitrary
three-dimensional configurations (``formation flying''), as will be required
by planned interferometers (LISA, Darwin).

Observing from space also brings with it disadvantages. After launch the
instruments are inaccessible, making maintenance and upgrades impossible. (The
serviceability of HST came at an enormous cost: A single Space Shuttle launch
alone cost about half a billion US dollars, so this will surely not be
repeated in a hurry.) Therefore the instruments, telescope and other systems
must be very robust, and this increases the cost. Even then, cosmic radiation
degrades electronic components -- in particular the CCDs -- and the optics,
degrading performance and complicating calibrations. Once instruments are
placed in distant orbits, e.g.\ the Earth--Sun L2 orbit, the available
bandwidth for data transfer is reduced (because for a given mass/size/cost of
satellite the power available is limited). This may restrict the amount of
science data which can be transmitted to the ground or it may impact the
observing strategy (both are the case for Gaia).

\section{Upcoming missions}

I now describe some upcoming missions which have a significant Galactic
astrophysics component. I limit myself to a selection of missions not yet
launched, I omitting projects in the very early planning stage.

\subsection{Herschel}

Herschel is a far infrared and sub-mm ESA observatory due for launch in 2008.
It comprises several imaging and spectroscopic instruments operating between
60 and 670\micron\ and is the only space facility dedicated to this part of
the spectrum. The key science objective of Herschel is the formation of stars
and galaxies.  Its major mode of operation will be deep, multi-band
photometric surveys (plus follow-up spectroscopy) to search for
proto-galaxies.  In this way it will investigate the formation and evolution
of galaxy bulges and elliptical galaxies during the first third of the present
age of the universe, determining how the galaxy luminosity function and star
formation rate has evolved with time. Herschel will also study the physics and
chemistry of the interstellar medium (both in our Galaxy and nearby galaxies)
and address the question of how stars form out of molecular clouds. While
surveys are expected to occupy much of its science time, Herschel is a
multi-user observatory, with two thirds of its time open via competitive
application.  Website:
http://www.rssd.esa.int/herschel
%\href{http://www.rssd.esa.int/herschel}{http://www.rssd.esa.int/herschel}

\subsection{JWST}

The James Webb Space Telescope will also be an infrared imaging and
spectroscopic facility but operating at shorter wavelengths than Herschel,
between 0.6 and 27\micron.  It is optimized for diffraction limited imaging in
the 2--5\micron\ region. With a 6.5m diameter primary mirror it will be able
to perform very deep imaging in ``pencil beam'' surveys (the field-of-view of
each instrument being just a few arcmin).  It has four instruments operating
over different wavelength ranges, three of which have spectroscopic modes with
resolving powers between 100 and 7000.  Although JWST was once conceived as
the successor to HST this is no longer the case, because it operates over a
different wavelength range (HST operated between 0.1--2.5\micron; JWST lacks
blue and UV sensitivity).  One of the main science objectives of JWST is to
study the early universe, in particular the epoch of the first stars and the
formation of the first galaxies. The other major theme concerns star and
planet formation in our own Galaxy and the study of exoplanetary systems. JWST
will be a multi-user observatory and so inevitably will be used for a very
wide range of topics.  Under the present plan, JWST should be launched in
around 2013.  Website:
http://www.jwst.nasa.gov/
%\href{http://www.jwst.nasa.gov/}{http://www.jwst.nasa.gov/}

\subsection{SIM PlanetQuest}

The Space Interferometry Mission plans to be the first long-baseline,
space-based interferometer. The original concept of synthetic imaging has been
dropped; the goal now is to perform astrometry to a limiting precision of
4\uas\ over a wide field (15\deg) and 1\uas\ over a narrow field (1\deg).
Unlike Hipparcos and Gaia, SIM is a pointed-mission, so the integration time
can be set according to the target magnitude and precision required.  Several
key programmes have been approved which relate mostly to planet finding and to
Galactic structure. The latter includes calibration of the stellar
Mass--Luminosity relation, measuring the distances and ages of globular
clusters and measuring the Galactic potential via stellar proper motions.
Dedicated observations taking about a quarter of the mission's time will be
used to measure reference frame objects to achieve absolute astrometry.  Many
of these science objectives will be covered ``automatically'' by Gaia (because
it is an all-sky survey), and with many more stars. However, SIM would be a
natural follow-up to Gaia, because, as a pointed mission, SIM could observe
selected sources more accurately. This complementarity with Gaia should be
taken into account when specifying the SIM science programme and reviewing
detailed proposals (a third of the observing time will be open to
competition).  At the time of writing (August 2006), the SIM (NASA) funding
situation is rather ``chaotic'' (to quote a senior scientist close to the
mission). It is unclear whether SIM will receive funding or when, but the best
estimate is for a launch no earlier than 2015.  Website:
http://sim.jpl.nasa.gov/
%\href{http://sim.jpl.nasa.gov/}{http://sim.jpl.nasa.gov/}

\subsection{Jasmine}

Jasmine is a project near infrared astrometry project.  The current design is
based on CCD detectors operating in the z-band with the Galactic
bulge as the target. Although its observing method does not permit it to
independently determine ``absolute'' proper motions, by observing stars with
proper motions accurately determined by Gaia an external calibration is
achieved.  If approved by the Japan Aerospace Exploration Agency (JAXA) when
proposed in 2010, it could launch by 2015. A related mission, Nano-Jasmine,
operating on the principle of Hipparcos and Gaia, will perform 1\,mas optical
astrometry using a 5\,cm telescope and should launch in 2008.  Jasmine is the
subject of several poster papers at this Joint Discussion and reported on in
these proceedings.  Website:
http://www.jasmine-galaxy.org/index.html
%\href{http://www.jasmine-galaxy.org/index.html}{http://www.jasmine-galaxy.org/index.html}

\section{Gaia}

\subsection{Gaia in a nutshell}

Gaia is an all sky astrometric and photometric survey complete 
to magnitude G=20 (V=20--22), which covers
10$^9$ stars, a million quasars and a few million galaxies. Gaia will achieve
an astrometric accuracy of 12--25\,\uas\ at G=15 (providing a distance
accuracy of 1--2\% at 1 kpc) and 100--300\,\uas\ at G=20.  These numbers are
also the approximate parallax accuracy in \uas\ and the proper motion accuracy
in \uas/year.  The range reflects the colour dependency: larger accuracy is
achieved for redder sources. Astrometry and photometry are done in a broad
(``white light'') band (G). Gaia will also measure radial velocities to a
precision of 1--15\,km/s for stars with V=17 via R=11\,500 resolution
spectroscopy around the CaII triplet (the ``Radial Velocity Spectrograph'',
RVS).  To characterize all sources (which are detected in real time), each is
observed via low dispersion prism spectrophotometry over 330--1000\,nm with a
dispersion between 3 and 30 nm/pixel. From this we will estimate the ``usual''
astrophysical parameters, \teff, \logg\ and \feh, but also [$\alpha$/Fe] and
the line-of-sight extinction to stars individually. (The radial velocity
spectrograph helps the parameter determination plus the detection of emission
lines and abundance anomalies.)

Gaia is a fully-funded ESA mission due for launch in late 2011. With a nominal
mission of five years and three years planned for post-mission processing, the
final catalogue will be available in about 2020.  It is the only large scale,
high-accuracy astrometry mission under construction.  For more information on
the satellite, science and data processing see
http://www.rssd.esa.int/Gaia
%\href{http://www.rssd.esa.int/Gaia}{http://www.rssd.esa.int/Gaia} 
and
\citep{turon05} (also available from the website).  In the rest of this
section I describe how Gaia will contribute to some important topics in
Galactic astronomy.

\subsection{Distances}

Distances are vital in every area of astronomy. We need them to convert 2D
angular positions to 3D spatial coordinates, allowing us to reveal the
internal structure of stellar clusters or map the location of the spiral arms,
for example.  Knowing the distance we can convert 2D (angular) proper motions
to physical velocities, and the apparent luminosity to the intrinsic
luminosity, a fundamental quantity in stellar structure and evolution studies.
Parallaxes are essentially the only method of direct distance determination,
and the only one which does not make assumptions about the target source. We
can measure parallaxes to virtually anything (not just, say, eclipsing
binaries) and virtually all other rungs in the distance ladder are ultimately
calibrated by them.

The impressive astrometric accuracy of Gaia is better illustrated when
convolved with a model of the Galaxy. This shows that Gaia will yield
distances with an accuracy of 1\% or better for 11 million stars. This
compares to fewer than 200 stars now with a parallax of this accuracy obtained
from Hipparcos, all of which lie within 10\,pc. Some 100\,000 stars will have a
distance accuracy better than 0.1\% and about 150 million better than 10\%.
Gaia goes far beyond anything we currently have in both accuracy {\em and}
statistics and hardly any field of astrophysics will remain untouched.

\subsection{Galactic structure and formation}\label{substructure}

One of the most important questions Gaia will address is that of how and when
the Galaxy formed. $\Lambda$CDM models of galaxy formation predict that
galaxies are built up by the hierarchical merger of smaller components
\citep{freeman02}.  Indeed, models predict that the halo is composed primarily
of the remains of mergers.  From extragalactic observations there is good
evidence for both the accretion of small components and for the merging of
similar-sized galaxies. Within our own Galaxy, recent surveys over the past
ten years -- 2MASS and SDSS in particular -- have found the fossils of past
and ongoing mergers in the halo and possibly also the disk of our Galaxy. They
have all been identified as spatial overdensities in two-dimensional (angular)
photometric maps of large areas of the sky. In some cases, distance measures
have been included by taking magnitude as a distance proxy \citep{belokurov06}
or by examining the 2D density of a limited range of spectral types (i.e.\ 
using a spectroscopic parallax of some tracers stars) \citep{yanny00}.  But
because merging satellites are disrupted by the Galactic potential and the
material spread out after several orbits, density maps are a limited means of
finding substructure.  Without an accurate distance the interpretation of 2D
maps is plagued by projection effects.  Even with perfect 3D maps, the
contrast against the background (including other streams) is often low
\citep{brown05}.  To improve this, we need 3D kinematics, i.e.\ radial
velocities and proper motions (combined with distances). In an axisymmetric
potential the component of angular momentum parallel to this axis ($L_z$) of
a merging satellite is an integral of motion.  In a static potential, the
energy ($E$) is also an integral of motion \citep{bt87}.  Thus while a merging
satellite could be well-mixed spatially, it would remain unperturbed in $(L_z,
E)$ space.  Of course, the Galactic potential is neither time-independent nor
perfectly axisymmetric and Gaia has measurement errors, but simulations have
demonstrated that Gaia will be able to detect numerous streams 10\,Gyr or more
(i.e.\ many orbits) after the start of disruption \citep{helmi00}.

Gaia will perform a 5D phase space survey over the whole sky, with the sixth
component -- radial velocity -- being available for stars brighter
than V=17. At this magnitude, spectral types A5III, AOV and K1III (which all
have M$_V$ $\simeq 1.0$) are seen at a distance of 16\,kpc (for zero
extinction). The corresponding proper motion accuracy is about 50\,\uas/yr, or
4\,km/s.  In converting proper motions to velocities the dominant source
of uncertainty (in this case) is the parallax error, which is about 100\% for
G=17 at 16\,kpc. In such cases, Gaia would rely on a ``spectroscopic''
parallax (calibrated, of course, with Gaia observations). In addition to the
5D or 6D phase space information, Gaia provides abundances and ages for
individual stars. Search for patterns in this even higher dimensional space
permits an even more sensitive (or reliable) search for substructure. To
properly exploit this data it will clearly be necessary to develop dynamic
models of the Galaxy with include stellar and chemical evolution.

\subsection{Dark matter}

Two distinct aspects of the Gaia mission permit us to study the mass and
distribution of dark matter in our Galaxy. First, from the 3D kinematics of
selected tracer stars, Gaia will map the total gravitational potential (dark
and bright) of our Galaxy, in particular the disk.  Second, from its
parallaxes and photometry Gaia will make a detailed and accurate measurement
of the stellar luminosity function. This may be converted to a (present-day)
stellar mass function via the Mass--Luminosity relation (see
section~\ref{ML_relation}).  From this we can infer a stellar mass
distribution. Subtracting this from the total mass distribution obtained from
the kinematics yields the dark matter distribution. This will be the first
time that the distribution of dark matter is accurately mapped on small length
scales (less than 1\,Mpc).

\subsection{Stellar structure, evolution and clusters}

Stellar luminosity is one of the most fundamental predictions of a stellar
model. Its measurement across a range of masses, ages and abundances is a
critical ingredient for testing and improving these models. In open and
globular clusters an accurate determination of luminosities and effective
temperatures (which Gaia also provides) gives us the HR diagram for different
stellar populations. (To derive an accurate luminosity we also need an
accurate estimate of the line-of-sight extinction. This will be obtained
star-by-star from the Gaia spectrophotometry.) We may then address fundamental
questions of stellar structure, such as the bulk Helium abundance (which is
not observable in the spectrum), convective overshooting and diffusion.  One
of the main uncertainties in the age estimation in clusters is accurately
locating the main sequence (for open clusters) or main sequence turn off (for
globular clusters). Gaia's accurate parallaxes and unbiased
(magnitude-limited) survey will greatly improve this.

In addition to using clusters as samples for refining stellar structure and
evolution, we can also study them as populations. Gaia will observe many
hundreds of clusters, allowing us to determine the (initial) mass function
into the brown dwarf regime and examine its dependence on parameters such as
metallicity, stellar density and environment. There are perhaps 70 open
clusters and star formation regions with 500pc. Gaia will provide distances to
better than 1\% individually for {\em all} stars brighter than G=15, and to
0.1\% or better for slightly brighter or nearer stars. This will permit us,
for the first time, to map the 3D spatial structure of many clusters, with a
depth accuracy as good as 0.5--1\,pc for clusters at 200\,pc. From the 3D
kinematics we can likewise study the internal dynamics of a cluster. Recall
that a proper motion of 1 mas/yr at a distance of 1\,kpc corresponds to a
speed of about 5\,km/s. A G=15 star will have its proper motion measured with
an accuracy of 20\,\uas/yr, corresponding to a speed uncertainty of 0.1\,km/s
at this distance (half this for a red star). The speed uncertainty varies
linearly with the distance for a fixed magnitude, so at 200\,pc the speed
uncertainty is just 20\,m/s. With this accuracy\footnote{This accuracy applies
  if the uncertainty in the transverse velocity is dominated by the proper
  motion error and not the distance error. This will be the
  case both for sufficiently nearby stars, or for clusters, in which a common
  distance can be assumed for the sake of proper motion to velocity
  conversion.}  we can measure the internal kinematics of the cluster and so
investigate the phenomena of mass segregation, low mass star evaporation and
the dispersion of clusters into the Galactic field.

Just as Gaia is an ideal tool for identifying the fossils of past mergers from
their phase space substructure (section~\ref{substructure}), so the 6D phase
space data plus astrophysical parameters for tens of millions of stars
will allow Gaia to detect new stellar clusters, associations or moving groups
based on their clustering in a suitable multi-dimensional parameter space. (It
can likewise confirm or refute the existence of controversial clusters.)  Work
is ongoing to develop and apply machine learning techniques to this problem.

\subsection{Stellar mass--luminosity relation}\label{ML_relation}

Gaia will detected many binary systems. These are found primarily via the
astrometry (nonlinear astrometric solutions which do not fit the standard
5-parameter model), but also as spectroscopic binaries in the RVS, eclipsing
binaries in the photometry, or as unresolved binaries from the detection of
two spectral energy distributions as part of the classification work.  For
those systems with orbital periods of about ten years or less, Gaia can solve
for the orbital elements and for the total mass of the system. If the
components of the system are spatially resolved then we may determine their
individual masses. Gaia furthermore measures accurate intrinsic
luminosities. Together these allow us to determine the stellar
Mass--Luminosity relation, and to do it with more stars and over a wider mass
range that has yet been performed.

\subsection{Data processing}\label{gaia_dp}

\begin{figure*}[t!]
\begin{center}
\resizebox{0.65\hsize}{!}{\includegraphics[clip=true]{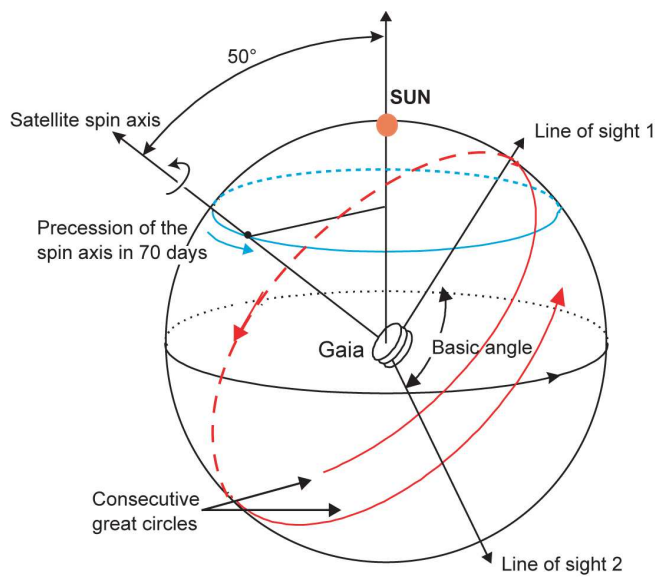}}
\caption{\footnotesize
The Gaia measurement principle and scanning law
\label{gaia_measurement_principle}
}
\end{center}
\end{figure*}

Gaia observes in two fields of view separated by a fixed {\em basic angle} of
about 100\deg. It continuously rotates around an axis perpendicular to the
plane formed by these two viewing directions. In a single rotation period of
six hours Gaia therefore observes (in each field) a great circle on the sky.
In addition, the rotation axis precesses (with a period of 70 days) such
that the rotation axis keeps a constant angle with respect to the Sun of
45\deg (see Fig.~\ref{gaia_measurement_principle}).
This two
axis motion, plus the orbit of Gaia about the Sun, defines the nominal
scanning law, i.e.\ how Gaia observes the sky.  The Gaia focal plane is
covered with just over 100 CCD detectors (with a total of 1\,Gpix) clocked and
read out in synchrony with the satellite rotation (``time-delayed
integration''). Over the course of its five-year mission\footnote{in
  which it really will ``boldly go where no-one has gone before'',
  scientifically at least.} the raw data product from each field of view
is essentially an image 7000\deg\ long by 0.7\deg\ wide (the field
width), in which each source appears an average of 40 times (it varies between
20 and 100 because the nominal scanning law does not give uniform sky
coverage). Each time the source is observed it will be displaced relative to
the others due to the orbit of Gaia (parallax, aberration), the source's
intrinsic motion (proper motion; orbital motion if it's a binary) and
gravitational light bending.

The goal of the astrometric data processing is to convert the 2D focal plane
positions in the two scanning strips into 2D angular positions, parallaxes and
proper motions. The steps in the reduction are: (1) Determine the source
centroid from the image. Because Gaia is continuously scanning, the position
is actually the time of the transit and its across-scan position. (2)
Knowledge of the calibration parameters for the CCDs and optics allows us to
transform the focal plane coordinates into field angles ($\xi$, $\eta$).  (3)
If the satellite attitude is known, the field angles may then be converted to
the proper directions.  (4) Correction of aberration and light bending (global
parameters) converts these field angles into astrometric positions.  By
observing simultaneously in two fields separated by the large basic angle,
Gaia is able to measure the {\em relative} positions between widely separated
objects. For any one object, the multiple scans are obtained at a wide range
of position angles and thus link this object to many different objects on the
sky. As we have measurements over the whole sky, this allows us to link all
objects and put all the astrometry on a common system.  From this multi-epoch
absolute astrometry we can derive, for each object, the five astrometric
parameters (mean position, parallax, proper motions) by fitting a
five parameter model to the 80 2D observations (160 measurements).

The mapping of the 2D on-board positions to absolute astrometry requires that
a large number of CCD/optical calibration and satellite attitude parameters
are known. However, given the very high accuracy which Gaia aims to achieve,
the nominal values of these parameters cannot be measured on ground to
sufficient accuracy.  Gaia must be {\em self-calibrating}. This is achieved in
the data processing with the {\em Astrometric Global Iterative Solution}
(AGIS). In this we solve for one set of parameters (e.g.\ the attitude
parameters) while holding the others (source, calibration, global) fixed. We
then iterate around the different sets of parameters until convergence. The
AGIS is run over a set of about 100 million stars (those with good fits to the
5-parameter model, selected iteratively), and involves several million
calibration parameters and tens of million of attitude parameters. A prototype
AGIS has been demonstrated to work on simulated data of 1.1 million stars
``observed'' for 5 years.  It is even possible to solve for some of the global
parameters, in particular the $\gamma$ parameter (which parametrizes the
accuracy of General Relativity (GR) in the Parametrized Post-Newtonian
formulation), and thus make an accurate test of GR via light bending from the
Sun and planets in the solar system. Once we have solved for all parameters,
we use these to derive source parameters for the (majority) of stars which do
not accurately fit the 5-parameter solution, including variable and binary
stars (or exoplanetary systems).

The heart of AGIS is a vast least-squares problem.  While conceptually
straight-forward, it is computationally intensive and to run in a reasonable
time must be split into parallel operations. The AGIS is, furthermore, just a
small part of the total data processing. There are many other operations
including: object cross-matching; photometric processing; extraction,
combination and calibration of the spectrophotometry and RVS spectra;
calibration of the CCDs in the face of radiation damage; attitude modelling;
GR effects; astrometric solutions for binary stars; object classification and
the estimation of astrophysical parameters; spectrophotometric variability
analysis; tracking and determining orbits for solar system objects. All of the
data processing -- from telemetry stream to final catalogue -- will be
undertaken by the Gaia Data Processing and Analysis Consortium (DPAC).
Following five years of studies, this consortium was formally started in mid
June 2006. It comprises some 250 members (not all full time) across 15
countries.

\subsection{Scientific exploitation of Gaia}

The Gaia data processing is a complex and challenging task. It demands the
development of novel methods for processing, analysing and mining the data.
Gaia will collect scientific data between 2012 and 2017, so the final
catalogue will only be available in about 2020 (although intermediate data
releases are planned). The final catalogue will be publicly available and will
contain positions, parallaxes, proper motions, photometry, spectroscopy,
radial velocities, classifications, stellar astrophysical parameters and
variability information.  This will give us the first opportunity to undertake
accurate, high-dimensional analyses (3D positions, 3D velocities, ages,
abundances etc.) of the Galaxy with large (tens/hundreds of millions) numbers
of stars. To properly exploit these data with the goal of understanding the
Galaxy's origin and evolution, much more sophisticated dynamical models of the
Galaxy will have to be developed (see James Binney's contribution to
\citep{turon05}). The sheer size and accuracy of the Gaia data will require a
fundamental rethinking of how we analyse and model data in this and many of
the other scientific areas.

Despite its impressive capabilities, Gaia does not do everything.  In crowded
areas of the sky not all stars can be downloaded (and the focal plane will
eventually saturate), so regions of the Galactic plane and the centres of
globular clusters will remain unexplored. RVS has a limiting magnitude of V=17
(ultimately for cost reasons), so most stars will not have radial velocities.
There are several complementary surveys which should be undertaken to fully
exploit the Gaia data. In particular, radial velocities for fainter stars
should be obtained with wide-field multi-object fibre spectrographs on
ground-based 4m or 8m class telescopes. Planned projects can realistically
observe a few million stars per year. Stellar parameters and individual
chemical abundances could be extracted from the same data.  More accurate
follow-up astrometry could be performed on selected faint targets or in more
crowded regions, either with SIM or the Large Binocular Telescope, for example
(and Pan-STARRS and LSST will provide proper motions for stars fainter than
G=20).  Infrared parallaxes in the Galactic plane (to see further through the
extinction) would extend Gaia's survey of the thin disk, spiral arms and star
forming regions. This may be done by Jasmine, but a deeper survey at, say,
2\,$\mu$m\ would be even better.  Gaia also will require some ground-based
observations for the calibration of its photometry, spectroscopy and stellar
parametrization algorithms. Ultimately, a full exploitation of the Gaia data
requires us to combine it with other data. We should start to think now about
what other data will be available in the timeframe 2015--2020, and if crucial
data will be lacking, to start making plans now to remedy this.

\section{In conclusion}

The next significant advance in understanding the formation, structure and
evolution of galaxies will come about from three lines of pursuit.  The first
is detailed astrometric and chemical abundance surveys of our own Galaxy (the
only galaxy where we can presently hope to make very detailed surveys). This
is addressed primarily by Gaia, but also by SIM and Jasmine if they fly.  Gaia
will deliver parallaxes accurate to 10--20\,\uas\ at G=15, yielding distances
better than 1\% for some ten million stars.  While our Galaxy retains fossils
of its evolution, these will only ever tell us part of the story, and then
only for one galaxy.  The second line of pursuit is the observation of
different galaxies at different stages of their life, i.e.\ at a range of
redshifts.  Several ground- and space-based surveys are already addressing
this, but the next generation of satellites, in particular JWST and Herschel
(operating in the infrared out to 670\,\micron) will focus much more on the
earliest epochs of galaxy formation (and extragalactic star formation) in the
high-redshift universe.  Space-based platforms are indispensable for accurate
parallaxes and for accessing most of the infrared, and so are essential for
the first two lines of pursuit.  Together they will significantly advance our
understanding of galaxy formation, dark matter, chemical evolution and stellar
structure and evolution (to know galaxies we must know stars).  The third line
is the development of powerful models and data analysis tools.  These are
essential for processing and then exploiting the Gaia data, but will also be
required to draw together knowledge obtained from our 'near-field'
cosmological studies (our Galaxy) and high-redshift galaxies.  Effort must be
invested into developing models and techniques with as much zeal as the
instrumentation and space platforms.

\begin{acknowledgements}
  I would like to acknowledge the efforts of the many people involved in Gaia,
  including the Gaia Science Team, ESA, EADS-Astrium and the Gaia Data
  Processing and Analysis Consortium.
\end{acknowledgements}

\bibliographystyle{aa}

\end{document}